\documentstyle[aps,preprint] {revtex}
\tightenlines
\def\bra{\langle}
\def\ket{\rangle}

\def\cap{\noindent}
\def\capp{\par\noindent}

\def\kk{{\cal K}} 
\def\beq{\begin{equation}}
\def\eeq{\end{equation}}
\def\bar{\begin{array}[b]}
\def\barc{\begin{array}}
\def\bart{\begin{array}[t]}
\def\ear{\end{array}}
\def\le#1{\label{eq:#1}}
\def\re#1{\ref{eq:#1}}

\begin{document}

\title { The Limiting Temperature of hot nuclei from microscopic 
Equation of State  }

\author{ M. BALDO$^{*}$, L. S. FERREIRA$^{**}$ and O.E. NICOTRA$^{* \dagger}$ }

\address{ $^{**}$ Centro de F\'\i sica das Int. Fund. and Dep. de F\'\i sica, 
Inst. Sup. Tecn. \\ Av. Rovisco Pais, 1096 Lisboa, Portugal}
\address { $^*$ Istituto Nazionale 
di Fisica Nucleare, Sezione di Catania
and Dipartimento di Fisica, \\   Universit\'a di Catania,
Via S. Sofia 64, I-95123 Catania, Italy } 
\address{ $^\dagger$ Dipartimento di Fisica, Universit\'a di Messina
\\ Salita Sperone 31, 98166, S. Agata, Messina }

\maketitle

\begin{abstract}   
The limiting temperature $T_{lim}$ of a series of nuclei is calculated 
employing a set of microscopic nuclear Equations of State (EoS). It is shown
that the value of $T_{lim}$ is sensitive to the nuclear matter Equation 
of State used. Comparison with the values extracted in recent phenomenological
analysis appears to favour a definite selection of EoS' s. 
On the basis of this phenomenological analysis, 
it seems therefore possible to check the microscopic calculations of the
nuclear EoS at finite temperature,
which is hardly accessible through other experimental informations.

\end{abstract}



\section {Introduction}

The knowledge of the Equation of State (EoS) of nuclear matter at finite 
temperature is one of the fundamental issues in nuclear physics.
Phenomenological information on the EoS can be obtained
from experimental data on heavy ion collisions at intermediate energies
and astrophysical observations on supernovae explosions and neutron stars. 
The nuclear matter EoS is believed to go through a liquid-gas
phase transition, as many theoretical calculations indicate 
\cite{FriedPand,big,Malfliet,Weber}. 
However, if this phase transition exists, 
does not possess a direct correspondence in finite nuclei, due
to the presence of the Coulomb and finite size effects. In particular,
the Coulomb interaction is of long range and strong enough to modify the nature
of the phase transition.  
Instead, it has been recognized by some authors
\cite{LevitBonche,SongSu}, that the nuclear EoS is related to 
the maximal temperature a nucleus can sustain before reaching mechanical
instability. This ``limiting temperature" $T_{lim}$ is mainly the maximal
temperature at which a nucleus can be observed.
\par
It has to be stressed that the reaction dynamics can 
prevent the formation of a true compound nucleus. The onset of 
incomplete fusion reactions can mask completely the possible
presence of fusion or nearly fusion processes. At higher energies,
 the heavy ion reaction 
can be fast enough that no (nearly) thermodynamical equilibrium can
be reached, as demanded in a genuine standard fusion-evaporation
reaction. However, combined theoretical and experimental analysis
\cite{natprc} indicate that a nearly equilibrium condition is
reached in properly selected multifragmentation heavy ion reactions 
at intermediate energy. 
The main experimental observation is the presence of a plateau in the so-called
''caloric curve", i.e. in the plot of temperature vs. total excitation energy
\cite{poch,wada,cibor,cussol}.
This behaviour was qualitatively predicted by the Copenhagen statistical model
\cite{bondorf} of nuclear multifragmentation. The relation between
multifragmentation processes and the nuclear EoS was extensively studied
by several authors within the statistical approach to heavy ion reaction 
at intermediate energy
\cite{bondorf2,bondorf3,botvina,gross,friedman,csernai,bondorf4}.
\par
In different experiments, various methods are used to extract from the data 
the values
of the temperature of the source which produces the observed fragments,
but a careful analysis of the data \cite{natprc} seems to indicate a 
satisfactory consistency of the results. In refs. \cite{natprc,natprl}
an extensive set of experimental data was analyzed and it was
shown that the temperature
at which the plateau starts is decreasing with increasing mass
of the residual nucleus which is supposed to undergo fragmentation. 
Both the values and the decreasing trend of this temperature 
turn out to be consistent with its interpretation as limiting 
temperature $T_{lim}$. According to this interpretation, at increasing 
excitation energy the point where the temperature plot deviates from
Fermi gas behaviour and the starting point of the plateau 
mark the critical point
for mechanical instability and the onset of the multifragmentation
regime. The corresponding value of the critical temperature can be calculated
within the droplet model, and indeed many estimates based on Skyrme forces
are in fairly good agreements with the values extracted from phenomenology
\cite{natprc,SongSu}. Moreover, the relation between nuclear 
matter critical temperature $T_c$ and $T_{lim}$ appears to be quite stable
and independent on the particular EoS and method used, which 
allows \cite{natprl} to estimate $T_c$ from the set of values of
$T_{lim}$.
\par
In general, one can expect that $T_{lim}$ is 
substantially smaller than the critical one, $T_c$. In fact, 
both the Coulomb repulsion and the lowering of the surface tension with 
increasing temperature tend to destabilize the nucleus with respect to infinite
nuclear matter. Since the surface tension goes to zero at the
critical temperature, $T_{lim}$ is reached much before $T_{c}$.
These predictions were checked in the seminal paper
of ref. \cite{LevitBonche}, as well as in further studies based
on macroscopic Skyrme forces \cite{SongSu}, for which a simple relationship was
established between $T_{lim}$ and $T_c$.
In ref. \cite{BaldoSong} it was shown,
however, that if microscopic EoS are used, the relationship between 
$T_{lim}$ and $T_c$ is not so simple and systematic as in the case of
Skyrme force EoS, and only a qualitative connection exists.\par
In this paper we consider the finite temperature EoS 
in the framework of microscopic non-relativistic and relativistic 
many-body theory  of nuclear matter and the corresponding
critical temperature. Then the limiting temperature for finite nuclei is
calculated on the basis of the corresponding EoS. The comparison with
 phenomenology shows the sensitivity of $T_{lim}$ to the microscopic EoS.
These results open the possibility of a direct check of the microscopic
theory of the nuclear matter EoS. Indeed, all the considered microscopic EoS
reproduce the empirical saturation point, but their behaviour
at finite temperature can be quite different. 

\section{The microscopic EoS}

Microscopic calculations of the nuclear EoS at finite temperature
are quite few. The variational calculation by Friedman and
Pandharipande\cite{FriedPand} was one of the first few semi-microscopic 
investigation
of the finite temperature EoS. The results predict
a liquid-gas phase transition, with
a critical temperature  $T_c = 18-20$ MeV. Later,
Brueckner calculations at finite temperature \cite{big} confirmed these
findings with very similar values of $T_c$.\par
The Van der Waals behaviour, which leads to the liquid-gas 
phase transition, was also found in the finite temperature
relativistic Dirac-Brueckner (DB) calculations of ref. \cite{Weber,Malfliet}. 
A liquid-gas phase transition was clearly observed, but at a much lower
value, $T_c \approx 10 MeV $. 
   It seems unlikely that such lower
critical temperature can be attributed to relativistic effects,
since the critical density is a fraction of the saturation one,
where relativistic effects are expected to play no role. It 
is more likely that this lower value of $T_{c}$ is due to the smaller value of
the effective mass, and we will present evidence of that later.\par
More recently, chiral perturbation theory at finite temperature
was used \cite{weise} to calculate the nuclear matter EoS, up to
three-loop level of approximation. The theory is a low density expansion,
and it appears appropriate to study the critical point,
where the density is a fraction of the saturation density. Again a
Van der Waals behaviour was found, with a critical temperature
$T_{c} \approx 25 MeV$.   \par
This set of nuclear matter EoS can be considered representative of 
the possible predictions from microscopic many-body theory.
Here in the sequel of this section we will remind briefly the 
non-relativistic Bloch and De Dominicis formalism, 
used in our calculations,
which is an extension to finite temperature of the 
Bethe-Brueckner-Goldstone (BBG) expansion. The formalism used in 
Dirac-Brueckner calculations at finite temperature is formally
very similar, as we will discuss later. For the chiral perturbation 
the formalism is of course quite different, and we refer the reader to the 
original paper \cite{weise}.\par
The finite temperature  Bloch and De Dominicis 
linked diagram expansion is based on  the Grand-canonical representation and
has the property to lead, in the zero temperature limit, to the 
BBG expansion of the ground state energy.
The grand canonical potential per particle $\omega$ is written as the 
sum of the 
unperturbed potential $\omega_0'$ and a correlation term $\Delta\omega$,
\beq
 \omega = \omega_0' + \Delta\omega 
\eeq
\cap
corresponding to the one-body grand canonical potential, and a 
power series expansion in the
interaction $H_1$ involving connected diagrams only, respectively.
The unperturbed potential is defined by,
\beq
\omega_0' = \omega_0 -\sum_k U_k n(k)
\eeq
\cap 
with $n(k)$ the finite temperature Fermi distribution, 
$\omega_0$ the grand canonical potential of the independent particle 
hamiltonian $H'_0$, and the summation over the single particle 
potential $U_k$ represents the first
potential insertion diagram \cite{big}. Therefore, 
$\omega_0'$ includes all one-body contributions and its explicit form reads

\begin{equation}
\bar{rl}
 \omega'_0 =
- {2\over \pi^2} \int_0^{+\infty} k^2 dk [& {1\over \beta}
 \log (1 + e^{-\beta(e_k \mu)})  \\
  &   \\
&+ U(k) n(k) ] 
\ear
\label{eq:omep}
\end{equation}
\cap
$\mu$ being the chemical potential, and

\begin{equation}
\bar{rl}
 \Delta\omega =
{2\over (2\pi)^3} \sum_{lSJT} \hat{J}^2 \hat{T}^2 
\int dq \int P^2 dP e^{-\beta(\overline{E}_{Pq} - 2\mu)} \\
&      \\
&  \cdot d(q,P) \arctan \left[ {\pi (q l | \kk^{SJT}(\overline{E}_{Pq}) |ql)
 q^2 \overline{Q}(q,P) \over d(q,P) } \right],

\ear
\label{eq:delta}
\end{equation}
\cap
where the density of state $d$ is given by,

\begin{equation}
d(q,P) = |{\partial \overline{E}_{qP}\over \partial q} | =
    |{2\hbar^2 q\over m} + {\partial \over \partial q} \overline{U}_{qP} |.
\label{eq:densta}
\end{equation}

The two-particle energy $\overline{E}_{qP}$, the Pauli operator
$\overline{Q}_{qP}$ and the potential felt by two particle $\overline{U}_{qP}$,
are all angle averaged quantities \cite{big}.
These angular averaging is
expected to be accurate, allowing us to make the contribution
of different channels additive, since then, only the diagonal
part of the finite temperature scattering matrix $K$ contributes. 
The quantum numbers $lSJT$ specify the two-body channel and
$\hat{A} = \sqrt{2A + 1}$. \par
The single particle potential and the two-body scattering matrix $K$ satisfy
the self-consistent equations

\begin{equation}
U({\bf k}_1) = \sum_{\sigma \tau} \sum_{{\bf k}_2} 
\bra k_1 k_2 \vert K(\omega) \vert k_1 k_2 \ket_A 
n(k_2).
\label{eq:ueq}
\end{equation}
\cap
and
\begin{equation}
\bar{rl}
 \bra\!\!\! &k_1 k_2 \vert K(\omega) \vert k_3 k_4 \ket\!\!=\!\!  \bra
 k_1 k_2 \vert v \vert k_3 k_4 \ket +\   \\
  &                 \\
 &\sum_{k_3' k_4'} 
\bra k_1 k_2 \vert v \vert k_3' k_4' \ket\ {n_{>}(k_3')n_{>}(k_4')
\over \omega - e} \
 \bra k_3' k_4' \vert K(\omega) \vert k_3 k_4 \ket.
\ear
\label{eq:kkk}
\end{equation}
\cap
In Eq. (\re{delta})
\begin{equation}
\bar{rl}
  \bra k_1\!\!\! &k_2 \vert \kk(\omega) \vert k_3 k_4 \ket 
  = \\
 &                    \\
  &(n_{>}(k_1)n_{>}(k_2)n_{>}(k_3)n_{>}(k_4))^{1\over 2} 
   \bra k_1 k_2 \vert K(\omega) \vert k_3 k_4 \ket 
\ear
\le{kmatt}
\end{equation}
\cap

\noindent
In all the previous equations $\omega = E_{k_1} + E_{k_2}$,
$e = E_{k_3'} + E_{k_4'}$,
with $E_k = \hbar^2 k^2/2m \, +\, U_k$.
Eq. (\re{kkk}) coincides with the Brueckner equation for the Brueckner 
$G$ matrix
at zero temperature, if the single particle occupation number $n(k)$
are taken at $T = 0$. At finite temperature $n(k)$ is a Fermi distribution.
In Eqs. (\ref{eq:kkk},\ref{eq:kmatt}) $n_{>}(k) = 1 - n(k)$.
It has to be noticed, that only the principal 
part has to be considered in the integration, thus
making $K$ a real matrix.
\par
Eqs. \re{ueq} and \re{kkk} have to be solved self-consistently 
for the single particle potential. 
For a given density and temperature we solve the self-consistent equations
along with the  Eq. (\re{ro}) for the chemical potential $\tilde{\mu}$, 
\cap\begin{equation}
\rho = \sum_k n(k) = \sum_k {1\over e^{\beta (E_k - \tilde{\mu})} + 1 }
\label{eq:ro}
\end{equation}
\cap
Then we obtain the grand canonical
potential $\omega$ from Eq. (\re{delta}).
Finally we extract the free energy per particle $f$
from the relation,
\beq
 f = \omega\rho + \tilde{\mu}.
\le{free}
\eeq
\capp
The pressure $p$ is calculated performing a numerical
derivative of $f$, i.e. $p \,=\, \rho^2 \partial f/\partial \rho$.
Notice that the chemical potential $\tilde{\mu}$ extracted from
Eq. (\re{ro}) does not coincide with the exact thermodynamical
chemical potential $\mu$ given by
\beq
\mu = {\partial F\over \partial N} = f + \rho ({\partial f\over 
\partial \rho } ) 
\le{tilde}
\eeq
\cap
which is the one actually adopted, in order to satisfy the Hugenholtz--Van
Hove theorem \cite{big}. \par
It turns out that \cite{big}  
the dominant diagrams in the expansion are the ones that correspond
to the zero temperature BBG diagrams, where
the temperature is introduced in the occupation numbers only,
represented by Fermi distributions, thus justifying this
commonly used procedure of naively introducing the temperature effect.  \par
The same prescription  has been used
in Dirac-Brueckner calculations. The formalism is therefore in principle 
very similar.

\section{The limiting temperature of finite nuclei}

Following
ref. \cite{LevitBonche} the limiting temperature can be evaluated within
the liquid drop model, which should be accurate enough for medium-heavy
nuclei. The nucleus is described in terms of a droplet surrounded
by a vapour, in thermal and mechanical equilibrium. 
In the model one adds to the droplet pressure and chemical
potential the contributions due to the Coulomb force and surface
tension, which are evaluated assuming a spherical droplet. These additional
terms read,
\beq
\bar{rl}
\delta P &= P_C + P_S = \left( {Z^2 e^2 \over 5 A}\rho - 2\alpha(T)\right)/R \\
\ &           \\
\delta \mu &= {6 Z^2 e^2\over 5 A R} 
\ear
\le{coul}
\eeq
\capp
where $R$ is the droplet radius $R = ({3A \over 4\pi \rho})^{1/3}$,
$\rho$ is the droplet density and for $\alpha(T) = \alpha_0 
(1 + {3\over 2} T/T_c) (1 - T/T_c)^{3/2} $, with $T_c = 20$ MeV the
nuclear matter critical temperature and the surface tension
at zero temperature $\alpha_0 =1.14 $ MeV fm$^{-2}$, obtained
from the semi-empirical mass formula.
The Coulomb interaction introduces and additional
positive pressure $P_C$ and a repulsive contribution to the
bulk chemical potential $\mu$, 
while the surface tension provides and additional
negative pressure term which tends to stabilize the system. At increasing
temperature the surface tension decreases and the system becomes
unstable against Coulomb dissociation. The simplest way to observe
the modifications introduced by these terms is to consider
the plot of the chemical potential as a function of pressure,
both for nuclear matter and for the droplet model.

The intersection
between the liquid and the vapour branches defines the coexistence
point in nuclear matter. The additional terms will only shift the liquid
branch, since the vapour is assumed to be uniform and uncharged,
leading to a new coexistence point. \par
This procedure was followed for the set of nuclear matter
EoS discussed in the previous section. 
At the lowest densities in the vapour region, needed in the 
calculations, the microscopic EoS was extended following ref. \cite{big}.

\section{Results and discussion}

To illustrate the procedure followed in the microscopic calculations
of EoS and $T_{lim}$ in the framework of many-body theory,
the nuclear matter free energy is reported in Fig. 1a  as a 
function of density for various temperature in the case of the Bonn B
potential \cite{Bonn}. The points indicates the actual microscopic 
calculations, the full lines the corresponding polynomial fits. The
figure illustrates the precision and stability of the numerical procedure.   
The three-body force, discussed in
\cite{big}, was included 
with adjusted parameters to reproduce the correct saturation point. 
From the free energy, by numerical derivative, one gets 
the pressure depicted in Fig. 1b. The critical temperature for
the liquid-gas phase transition is the lowest temperature for which
the isotherm is monotonic and the critical point is the corresponding
inflexion point on the isotherm. From Fig. 1b 
the critical temperature appears to be around $T_c \approx 18$ MeV,
slightly below the value obtained in ref. \cite{big} for the Argonne v$_{14}$
potential \cite{v14} ( $T_c \approx 20$ MeV ).
This shows that there is some sensitivity of $T_c$ on the NN interaction.
It has to be stressed  that the two EoS have very close saturation points. \par
As it is well known, the Dirac-Brueckner approach gives in general
a better saturation point than the conventional Brueckner calculations
(without three-body force).
It has been shown that this is mainly due to the modification of the
nucleon Dirac spinor inside nuclear matter, which can be described
by the contribution of the so-called Z-diagram \cite{BrownWeise},
corresponding to the virtual creation of a nucleon-antinucleon pair.
The Z-diagram can be viewed as a particular three-body force, which
is repulsive at all densities. 
The density dependence of this
contribution was studied in ref. \cite{BrownWeise} and was found
to be of the type $\Delta e = C \rho^{8/3}$, with the coefficient $C$
depending on the NN interaction. In ref. \cite{Bonn} it was found that
such a term can account very precisely for the difference between
the Dirac-Brueckner calculation and the corresponding 
non-relativistic Brueckner one.\par
Finite temperature Dirac-Brueckner calculations are quite few in the 
literature \cite{Malfliet,Weber}. Furthermore, for our analysis
we need the free energy as a function of density at small steps
of the temperature. Fortunately it is possible to estimate
accurately the temperature dependence of the free energy at a given density
by a simplified procedure, avoiding the complexity of the full
finite temperature Dirac-Brueckner calculations. Once the zero
temperature EoS is known, we assume that the free energy at $T \neq 0$
can be obtained by including the variations of both entropy and
internal energy of a free Fermi gas with the value of the effective
mass ( at $k = k_F$ ) equal to the one calculated at the same
density and at $T = 0$. In this way one neglects the variation with
temperature of the effective mass and of the interaction energy.
Both these variations turn out to be small at the Brueckner level
\cite{big}, and indeed the same procedure applied to to non-relativistic
Brueckner calculations give excellent agreement with the full calculations
\cite{big}. \par
We applied this procedure to the EoS of ref. \cite{Malfliet}, by 
fitting the Dirac-Brueckner EoS at $T = 0$ and calculating the 
free energy at finite temperature from the corresponding effective mass. 
At variance with the previous calculations of ref. \cite{big}, we preferred
here to fit directly the EoS at zero temperature instead of applying
the relativistic correction due to Z-diagram mentioned above. This
should avoid any possible bias from the NN interaction. In any case,
the final results are quite similar to the previous calculations.    
We found a critical 
temperature $ T_c \approx 12$ MeV, in comparison with the value of 10 MeV
reported in ref. \cite{Malfliet}. This reasonable agreement
is a further check of the simplified procedure adopted.
Since the limiting temperature $ T_{lim} $ is expected to be a small fraction
of the critical temperature $T_c$, the error introduced by the 
simplified procedure can be
considered small enough for an accurate treatment of the Dirac-Brueckner
case. 
\par
In DB calculations the single particle energy $E_k$ is written as \cite{Bonn}

\beq
E_k = \sqrt{M^{*2} \, +\, k^2} + U_V \ \ \ , \ \ \ M^* \,=\, M + U_S
\eeq

\noindent
where $U_S$ and $U_V$ are the scalar and vector single particle
potentials respectively. In the non-relativistic limit the square root is
expanded in power of $k/M^*$. If one neglects
the momentum dependence of the scalar and vector
potentials, $M^*$ can be identified with the non-relativistic 
effective mass to be used in the finite temperature calculations
for the Fermi gas model. In the region of the liquid-gas phase 
transition the non-relativistic expansion is fully justified.
This is equivalent to a parabolic approximation for the single
particle energy. This procedure results in values of the effective mass
which are substantially smaller than in the conventional non-relativistic 
Brueckner calculations \cite{Bonn}, where 
no parabolic approximation for the single particle
potential is used \cite{Fiasconaro}. 
\par
For the EoS calculated within chiral perturbation theory, 
all the expressions are semi-analytical and the whole procedure is
much simpler.
\par
Plots of the chemical potential as a function of pressure
for nuclear matter are reported in Fig. 2 . 
The intersection
between the liquid and the vapour branches defines the coexistence
point in nuclear matter. Increasing the temperature, the curve 
shrinks and should collapse to a point at $T_c$, which can be thus determined
in this way. The values extracted along this procedure are in good
agreement with the values obtained from the plot of pressure vs. density,
Fig. 1b. This illustrates the consistency and precision of the numerical
procedure.
\par
For the droplet model,
including the corrections of Eq. (\re{coul}),
the new liquid branch, indicated by the dashed lines in Fig. 2,
shows a shift with respect to nuclear matter. At low enough
temperature an intersection between the liquid and vapour branches still
occurs, which corresponds to the coexistence point between the liquid
droplet and the nuclear matter vapour and assures that the
droplet is stable. Increasing the temperature, the curve 
shrinks and well below $T_c$
it is possible to find a temperature for which 
the intersection between the liquid droplet and the vapour branches
just disappears, as indeed reported in Fig. 2. This determines $T_{lim}$.
\par
The droplet-vapour coexistent point, and consequently $T_{lim}$, depends on 
the mass and charge of the system. 
\par
Fig. 3 summarizes the results of the calculations, in comparison with
the data obtained from the phenomenological analysis \cite{natprc,natprl}.
For completeness and for sake of comparison, 
also the results for the Av$_{14}$ potential of ref. \cite{big} is reported.
The calculated values of the limiting temperature $T_{lim}$,
for the considered set of microscopic nuclear matter EoS,
show an overall trend which clearly reflect the corresponding
trend for the critical temperature $T_{c}$ of each EoS. Smaller
values of $T_{c}$ results in a smaller value of $T_{lim}$.
\par
The ratio between $T_{lim}$ and $T_{c}$ 
for Skyrme forces was extensively studied in ref. \cite{natprl}. It was found
that this ratio is close to $1/3$ with a small dispersion. The
microscopic EoS analyzed in Fig. 3 give values which follow closely 
this value, except the Dirac-Brueckner case, which gives a value closer to 
$1/4$ This could be attributed to the approximate procedure we used for
this EoS, but in any case a value of $1/3$ would not alter the trend
reported in Fig. 3. 

\par
More importantly, the comparison of the values of $T_{lim}$ 
from microscopic EoS with the phenomenological values
 emphasizes the sensitivity of $T_{lim}$ to the EoS. This 
comparison appears as a crucial test for any microscopic EoS.
The EoS from ref. \cite{weise}, 
as noticed by the authors, produces a too large value of the 
nucleon effective mass, and this is probably the reason of the
too high value of $T_{c}$. In fact, a large effective mass
reduces the increase with temperature of the kinetic energy  and
therefore of the free energy.\par 
On the contrary, the DB results seem to
indicate that the corresponding EoS has a too small $T_{c}$.
Notice that this would be very difficult to verify with other phenomenological
analysis. The reason for such a small value of $T_{c}$, and
therefore of a too small value of $T_{lim}$, can be attributed again to the 
value of the effective mass, which is smaller than in 
the non-relativistic case. However, other characteristic of the EoS
could play a role, like the values of the chemical potential or
of the compressibility at low density (i.e. in the gas phase).\par
The non-relativistic BHF results appear to agree quite closely with the
phenomenological values. Some dependence on the NN interaction 
is present, but this uncertainty is within the phenomenological
uncertainty. Therefore, phenomenology appears to favour this
set of EoS. These results also support the interpretation of $T_{lim}$
as the temperature for the mechanical instability and the onset of
the multifragmentation regime. 

\section*{Acknowledgments}
We thanks very much Dr. N. Kaiser for his kindness in providing us 
an extended numerical
table of the Equation of State developed in ref. \cite{weise}. 
\par
One of us ( O. E. N. ) expresses many thanks for the kind hospitality
during his stay at the Centro de F\'\i sica das Int. Fund. 
in Lisbon, where part of this work has been developed.

\newpage
\par\noindent
{\bf Figure captions}
\vskip 0.4 cm
\noindent
Fig. 1a - Free energy per particle as a function of Fermi momentum at
different temperatures for the Bonn potential. From top to bottom 
the different curves 
correspond to temperatures $T= 2, 8, 12, 16, 20, 24, 28$ MeV.
The points represent the results of the Brueckner-Hartree-Fock calculations
at finite temperature, the curves are the corresponding polynomial fits.
\vskip 0.2 cm
\noindent
Fig. 1a - Isotherms of pressure vs. Fermi momentum corresponding
to the free energy plots of Fig. 1a. The sequence of temperatures
is the same as in Fig. 1a (from bottom to top).
\vskip 0.2 cm
\noindent
Fig. 2 - Chemical potential vs. pressure for the Bonn potential
from the Brueckner-Hartree-Fock calculations of Figs. 1a,1b (full line)
at a given temperature. The dotted line indicates the corresponding
plot for the nucleus $^{208}Pb$. At this temperature the nucleus
starts to be unstable, see the text for details.
\vskip 0.2 cm
\noindent
Fig. 3 - Limiting temperatures as a function of mass numbers for different
Equation of State in comparison with the phenomenological values (open
squares with error bars).

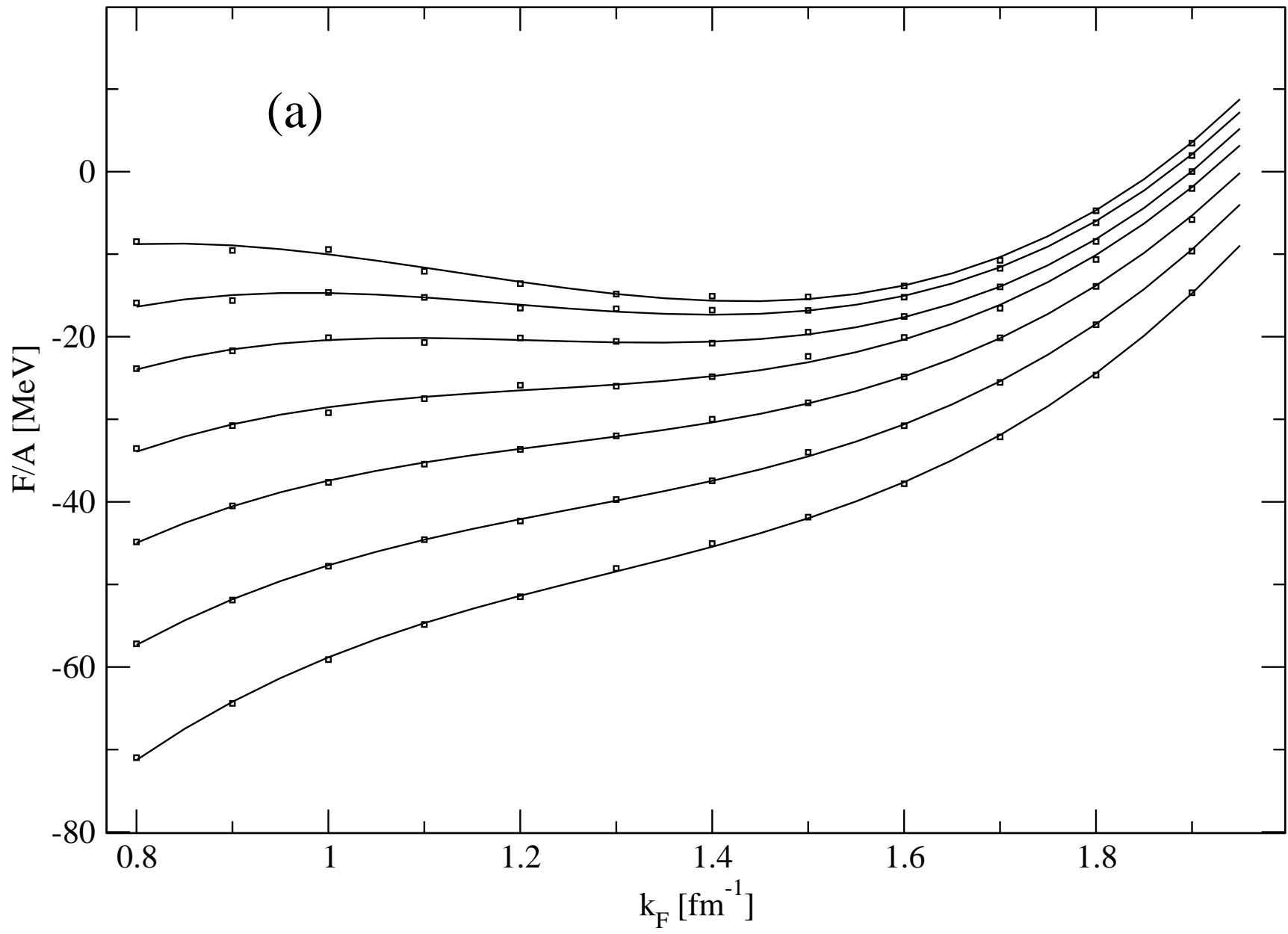
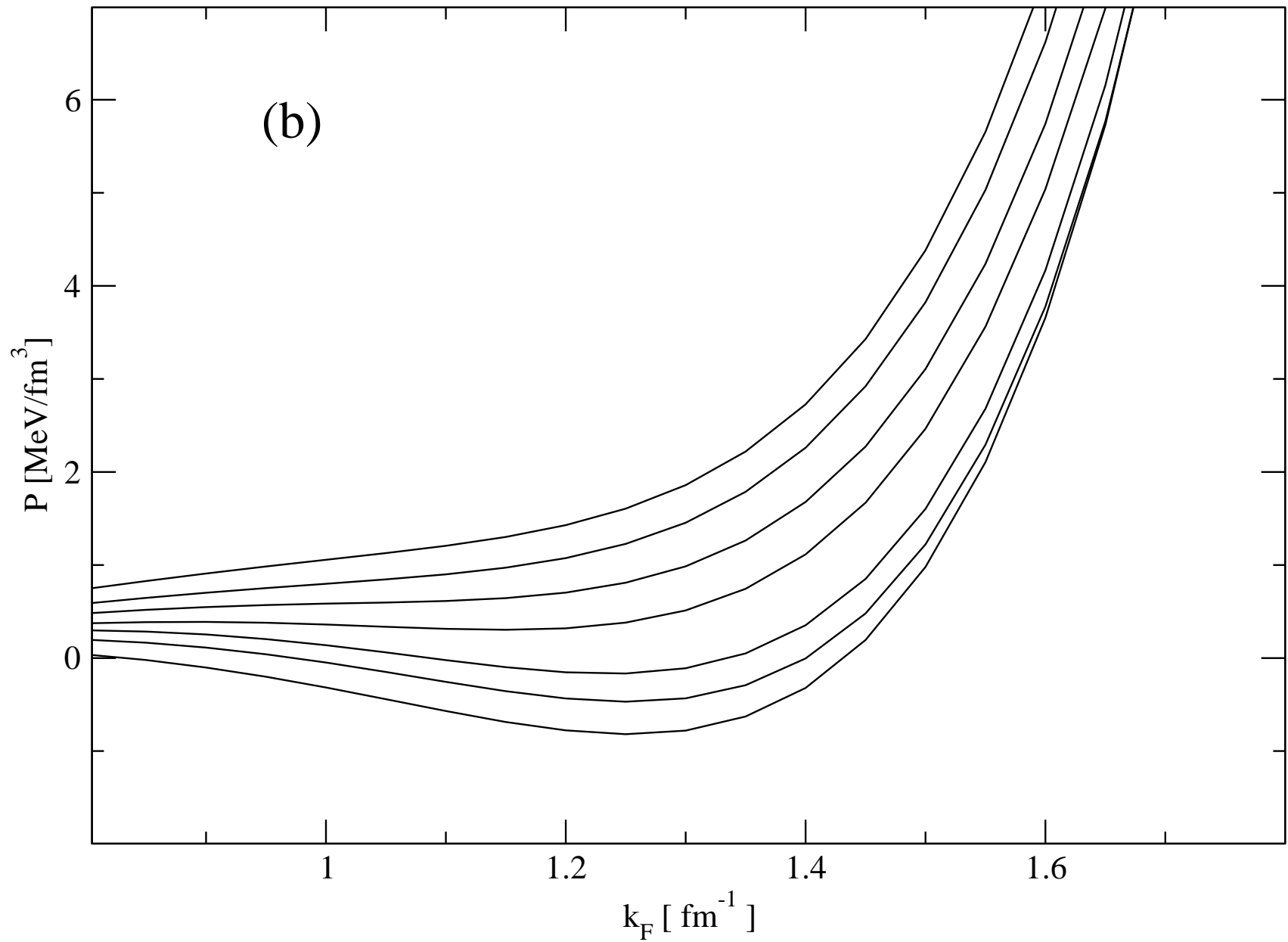
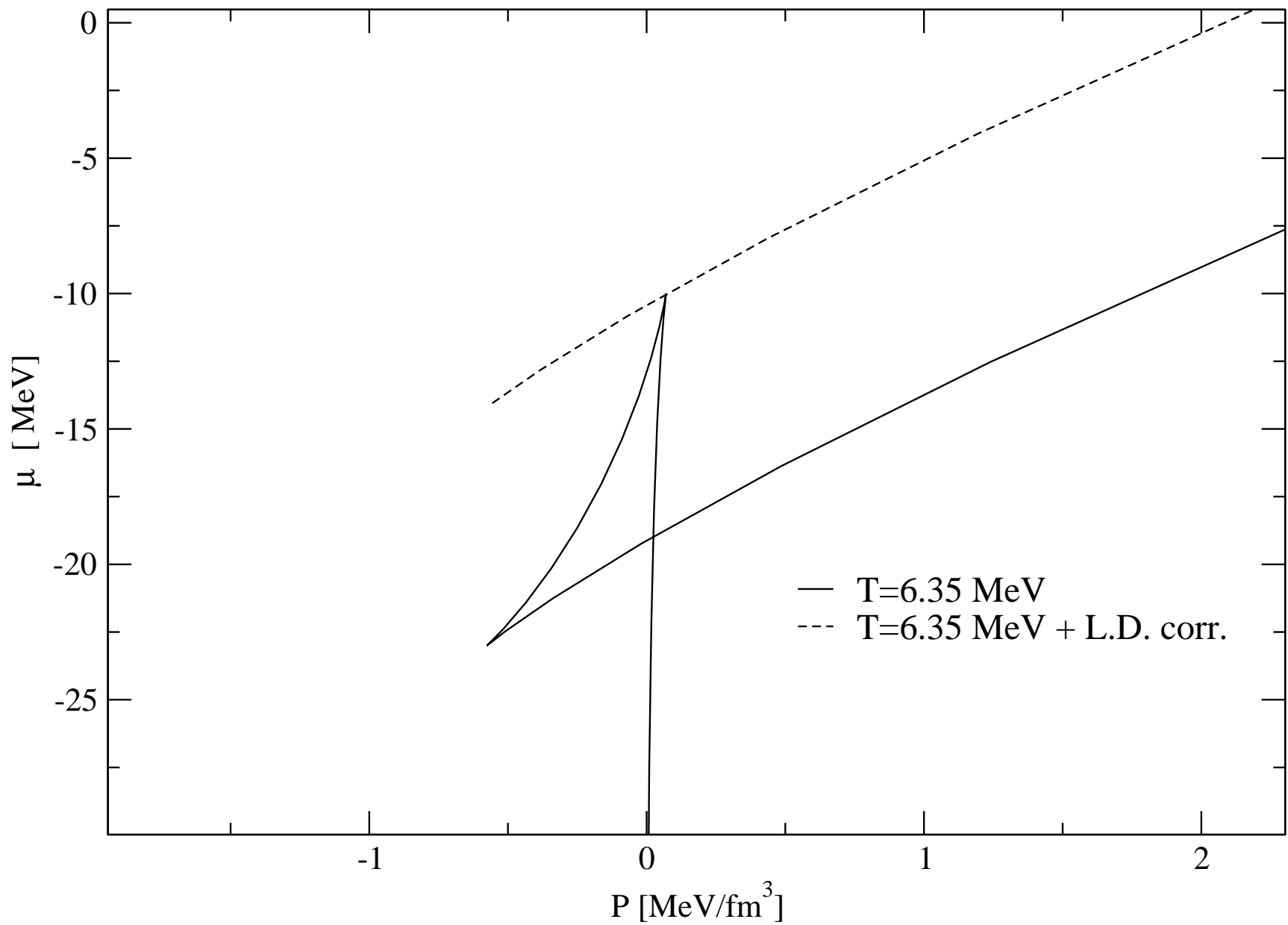
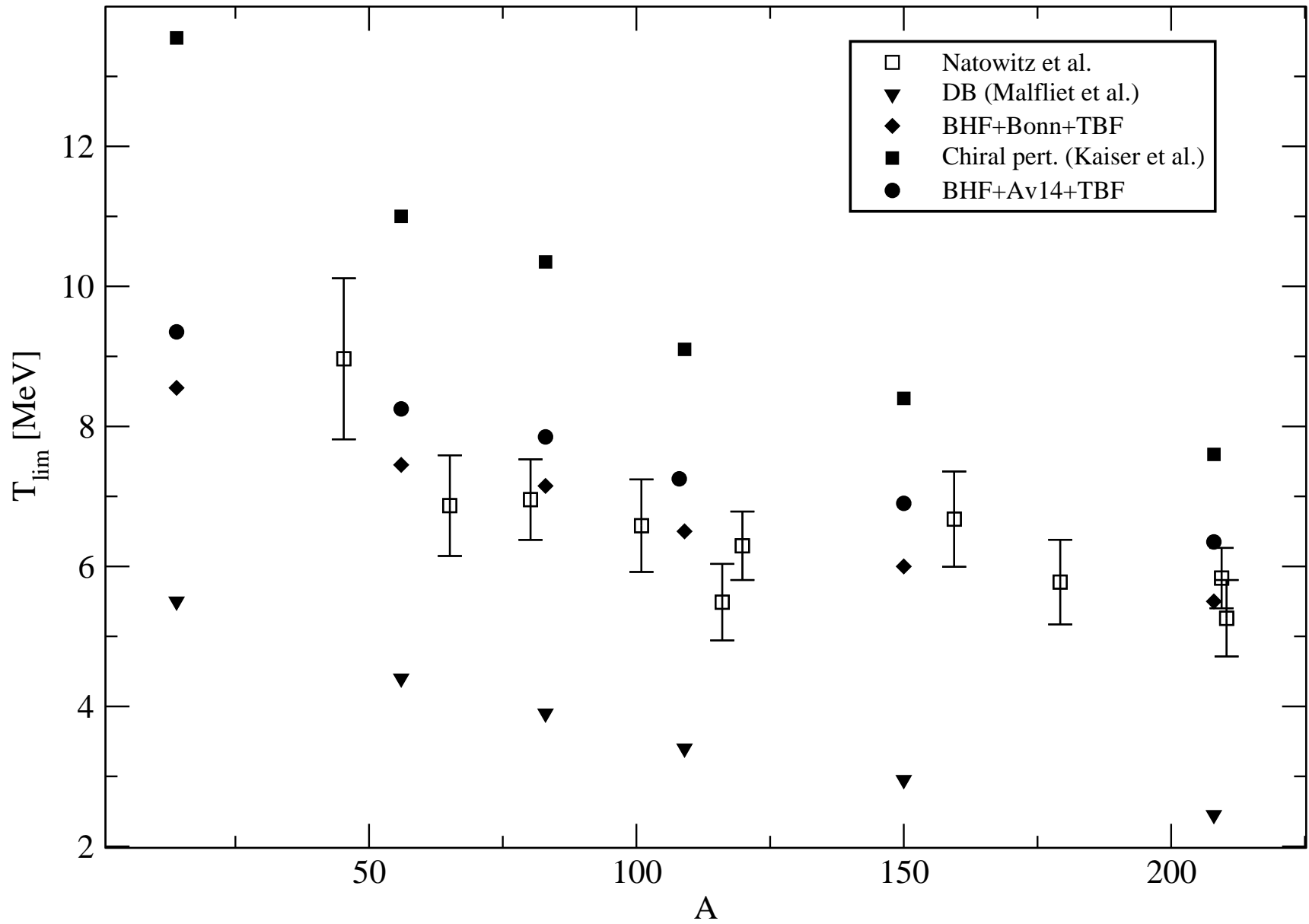
\end{document}